\documentclass{emulateapj}

\newcommand{\Kepler}{{\it Kepler}}

\usepackage{hyperref}
\usepackage{breakurl}
\usepackage{amsmath}
\slugcomment{}

\shorttitle{K2 Campaign 0 Light Curves}
\shortauthors{Vanderburg}

\begin{document}


\title{Reduced Light Curves from Campaign 0 of the K2 Mission}


\author{Andrew Vanderburg\altaffilmark{1,2}}
\affil{Harvard--Smithsonian Center for Astrophysics, 60 Garden St., Cambridge, MA 02138}

\altaffiltext{1}{avanderburg@cfa.harvard.edu}
\altaffiltext{2}{NSF Graduate Research Fellow}


\begin{abstract}
After the failure of two reaction wheels and the end of its original mission, the \Kepler\ spacecraft has begun observing stars in new fields along the ecliptic plane in its extended K2 mission. Although K2 promises to deliver high precision photometric light curves for thousands of new targets across the sky, the K2 pipeline is not yet delivering light curves to users, and photometric data from K2 is dominated by systematic effects due to the spacecraft's worsened pointing control. We present reduced light curves for 7743 targets proposed by the community for observations during Campaign 0 of the K2 mission. We extract light curves from target pixel files and correct for the motion of the spacecraft using a modified version of the technique presented in Vanderburg \& Johnson (2014). We release the data for the community in the form of both downloadable light curves and a simple web interface, available at \url{https://www.cfa.harvard.edu/~avanderb/k2.html}. This ArXiv only report is meant to serve as data release notes -- for a refereed description of the technique, please refer to Vanderburg \& Johnson (2014). 

\end{abstract}

%\keywords{Data Analysis and Techniques}

\section{Introduction}

After four years in space and the discovery of several thousand planet candidates, the \Kepler\ mission ended with the failure of the spacecraft's second reaction wheel in May of 2013. Without at least three functional reaction wheels, \Kepler\ could no longer point precisely at its original field. Through clever engineering, Ball Aerospace and the \Kepler\ team devised a way to stabilize the spacecraft against Solar radiation pressure by pointing away from the Sun in the plane of \Kepler's orbit and balancing \Kepler's solar panels against the Solar photons. This strategy has been successfully implemented in the K2 mission \citep{howell}.

The original \Kepler\ mission achieved high photometric precision in part because of \Kepler's ultra-stable pointing enabled by its reaction wheels. The loss of two reaction wheels hurt \Kepler's photometric precision, degrading the quality of the photometry by roughly a factor of four (Howell et al. 2014). Recently, using data from a 9 day engineering test conducted in February of 2014, \citet{vj14}, hereafter VJ14, showed that an improvement could be made to K2 photometric precision by decorrelating photometric light curves with the motion of the spacecraft. By measuring an empirical ``flat field'' and removing it, VJ14 showed that the quality of K2 photometry could be corrected to within a factor of 2 of the original \Kepler\ performance. 

In this report, we apply a similar procedure to reduce light curves for nearly 8000 targets observed by K2 during Campaign 0, a final shakedown before the beginning of real K2 science operations. We produce light curves from K2 target pixel files, and decorrelate the motion of the spacecraft from the photometry. We release these light curves to the community as the latest of several resources to enable widespread use of K2 data \citep{k2catalog, k2varcat}. 

\section{Data Acquisition}\label{dataacquisition}
Although Campaign 0 was supposed to be a full length shakedown test of the K2 operating mode, \Kepler\ was not able to attain fine guiding control for the first half of the test. After a mid-campaign data download, \Kepler\ managed to achieve fine pointing control for approximately 35 days during May of 2014. After basic processing by the \Kepler/K2 pipeline, target pixel files for 7743 targets and a ``super-stamp'' encompassing the clusters M35 and NGC 2158 were released on the Mikulski Archive for Space Telescopes (MAST) in early September of 2014. These target pixel files were subsequently re-released (as Data Release 2) in November of 2014, correcting errors in astrometric information in the target pixel file headers. We downloaded and performed our analysis on target pixel files from Data Release 2, taking advantage of the accurate astrometric information.

\section{Data Processing}\label{processing}
\begin{figure*}
\epsscale{1}
  \begin{center}
      \leavevmode
\plotone{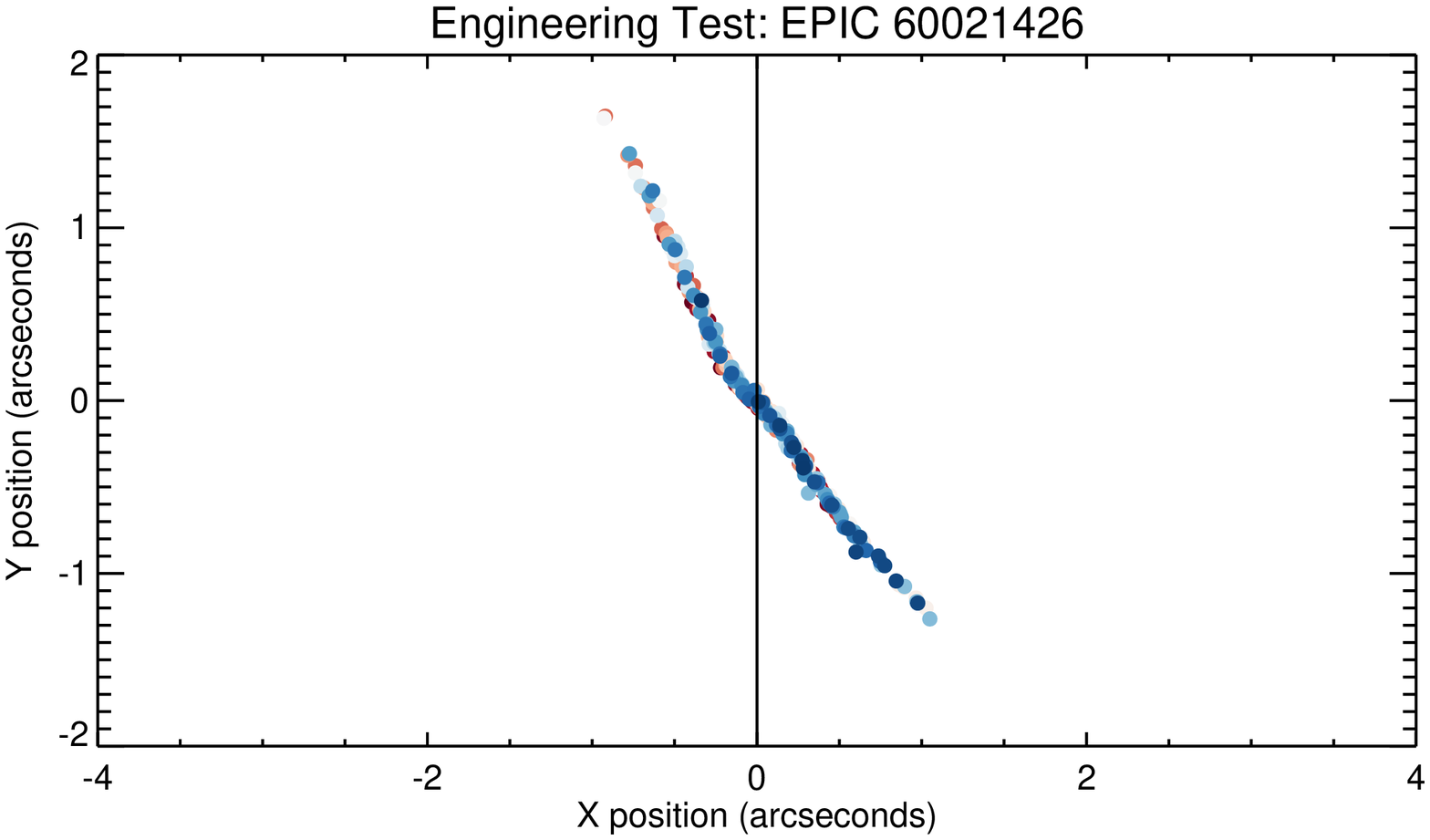}
\plotone{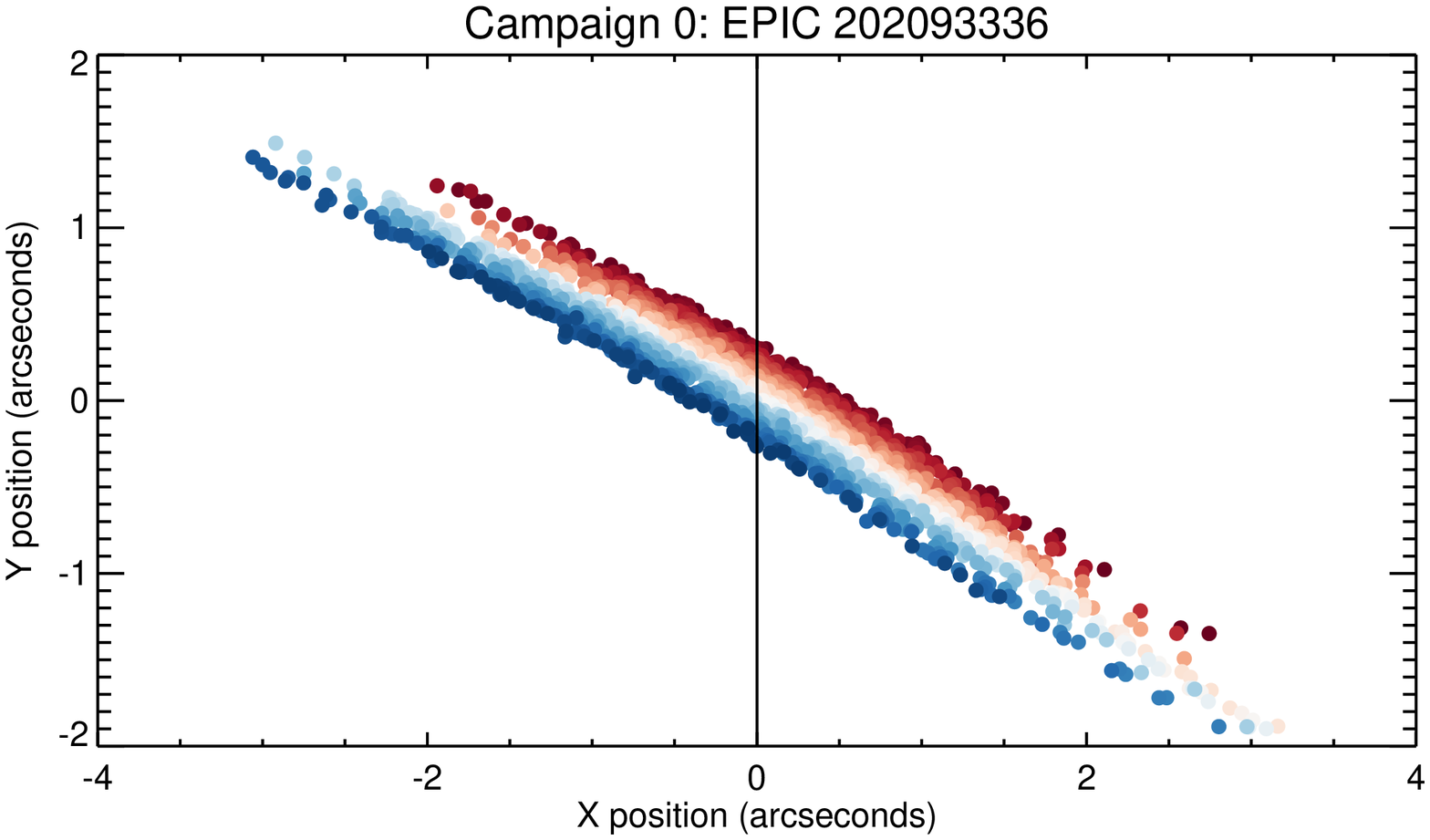}
\caption{Top: Image centroid positions from the engineering test. The color of the points denotes the time of the observation. Over the course of 6.5 days, \Kepler's pointing jitter traced out a one dimensional path along the detector, and there are no evident correlations with the time of the observation. Bottom: Image centroid positions from Campaign 0. The path across the detector is now no longer strictly one dimensional, and the evolving color transverse to the back and forth motion is evidence that the path is drifting with time. For both targets, each box is the size of one Kepler pixel.} \label{centroids}
\end{center}
\end{figure*}

We processed the Campaign 0 target pixel files using a descendant of the algorithm of VJ14. In brief, VJ14 performed aperture photometry on the K2 target pixel files and measured image centroid positions as a proxy for the motion of the spacecraft. They then excluded thruster firing events and other data points marked by the \Kepler\ pipeline as having suboptimal quality. Critically, they measured a one-dimensional correlation between the flux measured from each target and the position of the image on the detector. After measuring the correlation (which they called the ``self-flat-field'' or SFF), they fit it with a piecewise linear function and divided it from the raw light curve, yielding higher quality photometry. 

\begin{figure*}[t!]
\epsscale{1}
  \begin{center}
      \leavevmode
\plotone{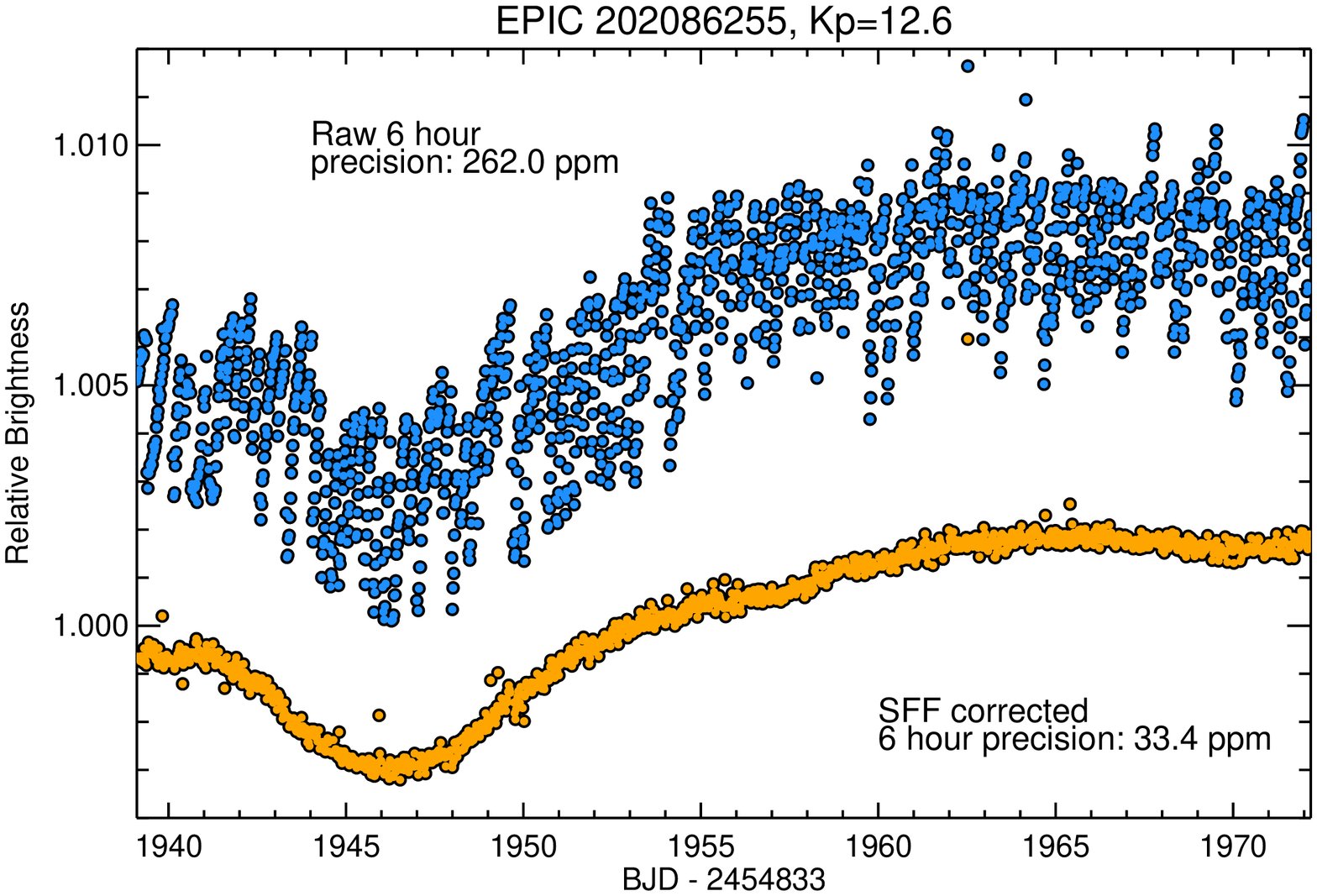}
\caption{Raw light curve (top) and SFF corrected light curve (bottom). The SFF correction significantly improves photometric precision while retaining long timescale stellar variability.} \label{sffcorrect}
\end{center}
\end{figure*}

We made several modifications to the procedure of VJ14 in processing the Campaign 0 light curves, which we list here: 
\begin{enumerate}
\item We broke up the 35 day campaign into three separate ``divisions'', and performed the SFF correction on each division separately.  VJ14 performed the decorrelation assuming the \Kepler's motion was a one dimensional path along the detector, a valid assumption over the 6.5 days of data used in their analysis. Over the course of the 35 days of Campaign 0 data, however, this assumption broke down due to a slow drift in the image centroid positions over the length of the campaign (see Figure \ref{centroids}). We were able to circumvent this problem by performing the SFF correction on shorter sections of data when the one dimensional approximation was still valid. We were able to recover long period stellar variability by iteratively fitting a basis spline (B-spline) to the low frequency variations in the entire light curve, and correcting each division separately, similar to the iterative procedure in VJ14. 

\item Instead of measuring the image centroid position for each star and using that to decorrelate   \Kepler's motion, we chose one bright but unsaturated star (EPIC 202093336), and used the centroids measured for that star for all other targets. This increased the robustness of the decorrelation, particularly for faint stars, ones with high background flux levels, and stars with nearby companions.

\item We extracted photometry from 20 different apertures and chose the aperture with the best photometric precision, as opposed to VJ14, who extracted photometry from two different apertures and chose the one with the best photometric precision. In particular, we extracted photometry for 10 circular apertures of different sizes and 10 apertures defined by the  \Kepler\ Pixel Response Function \citep[PRF,][]{keplerprf}, similar to the PRF defined apertures in VJ14, but choosing pixels with modeled flux greater than 10 different levels.  We selected the best aperture by selecting the fully processed light curve with the best 6 hour photometric precision as defined by VJ14. For stars fainter than Kp = 13.5, we limited the maximum size of the aperture to prevent brighter stars from dominating the flux in a large aperture. This agnostic approach to aperture selection improved the precision for faint stars by making it possible to select the smallest possible aperture while still performing a high quality SFF correction. Also unlike VJ14, we were able to take advantage of astrometric information in the K2 target pixel files to pinpoint the location of the right target. 

\item The motion of the stars on the detector is larger in Campaign 0 than it was in the engineering test (as is evident in Figure \ref{centroids}, so we compensated by using a 25 point piecewise linear fit to the self flat field, as opposed to the 15 point piecewise linear fit used by VJ14.  
\end{enumerate}

We show an example light curve before and after SFF processing in Figure \ref{sffcorrect}. The SFF processing improves photometric precision (as defined by VJ14) by a factor of 8 in this case.

\section{Characteristics of Campaign 0 Photometry}\label{characteristics}
\subsection{Photometric Precision}

\begin{figure*}[t!]
\epsscale{1}
  \begin{center}
      \leavevmode
\plotone{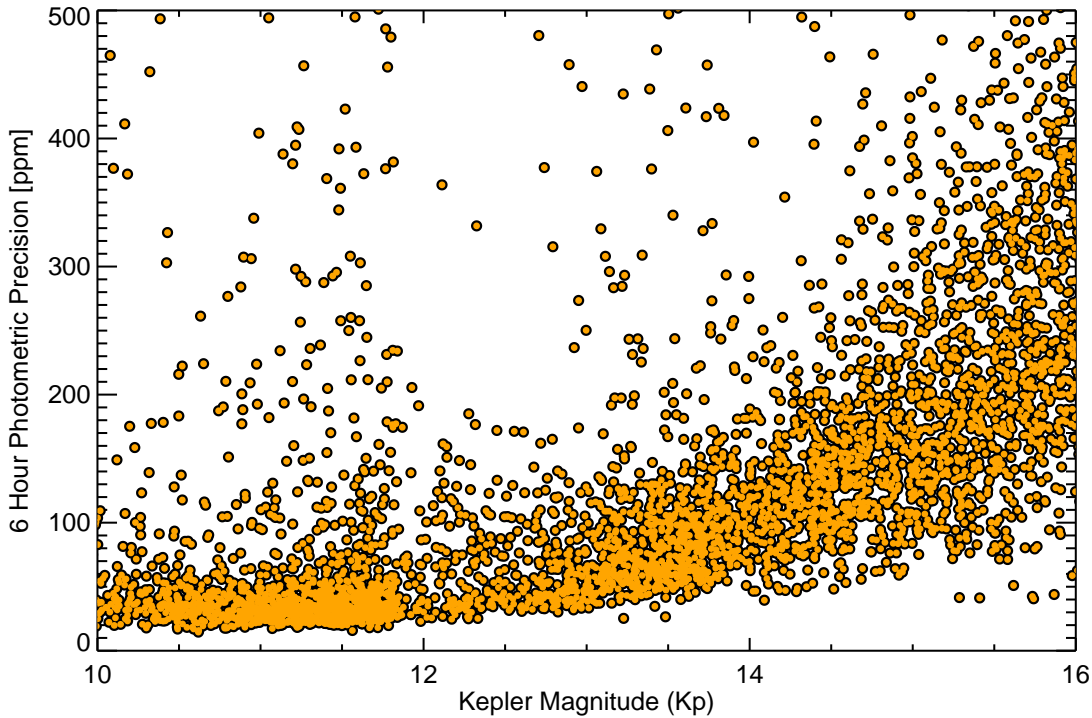}
\caption{Six hour photometric precision as a function of Kepler magnitude. Here, we plot only dwarf stars from the Guest Observer proposals GO0111 and GO0119. For visual clarity, we added a random scatter of 0.03 magnitudes to the Kepler magnitudes, which were reported rounded to the nearest 0.1 magnitude.  } \label{kpcdpp}
\end{center}
\end{figure*}

We extracted and corrected light curves for 7743 targets observed by K2 in Campaign 0. We did not extract light curves for targets in the super stamp. Like VJ14, we found that the SFF algorithm works best for dwarf stars, and can produce poor results for stars with rapid or high levels of photometric variability.

We assessed the photometric precision achieved by K2 during Campaign 0 and compared it to both the original \Kepler\ precision and the photometric precision achieved by K2 during the Engineering Test, reported by VJ14. Like VJ14, we measured photometric precision based on observations of cool dwarf stars, which have less astrophysical noise than giants and hot stars. We isolated dwarf stars observed by K2 during Campaign 0 by selecting stars proposed by two Guest Observer proposals: GO-0111 (PI Sanchis-Ojeda) and GO-0119 (PI Montet). These two proposals focus on exoplanet detection and therefore proposed almost exclusively dwarf stars. These were also the two largest proposals in terms of targets awarded, with 5028 targets between the two. We summarize the precision of C0 data compared to the precision of \Kepler\ data and K2 data from the engineering test in Table \ref{precisioncomp}. We plot the photometric precision of our targets versus their \Kepler\ band magnitude in Figure \ref{kpcdpp}.

We find that for faint stars (in particular stars with \Kepler\ band magnitudes between 14 and 15), we achieved better photometric precision than VJ14 in the Engineering Test. We attribute this to our more flexible aperture selection, which helps to limit background noise for faint, background dominated targets. For bright stars, we measure worse photometric precision than reported in VJ14. One possible explanation for the decreased precision compared to the engineering test is that Campaign 0 observed stars near the Milky Way disk, while the engineering test field pointed out of the galaxy. Observing targets close to the galaxy poses a challenge for \Kepler\ due to its large pixels and relatively poor (focus limited) spatial resolution. Source crowding could add additional noise, both by contamination with noisy sources and by increasing systematics due to sources entering and exiting the aperture due to K2's pointing jitter. Moreover, observing near the disk of the galaxy increases the likelihood of giant stars being accidentally included. Giant stars have higher levels of short timescale photometric variability than dwarf stars \citep[e.g.][]{bastien} and can increase the measured noise.

\begin{table}[t!]
\begin{center}

\caption{Median 6 Hour Photometric Precision}
\begin{tabular}{llll}
\hline\hline
$K_{p}$ & ET K2 & C0 K2 & \Kepler\ \\
\hline
10-11            & 31    & 38   & 18         \\
11-12            & 33    & 37   & 22         \\
12-13            & 40    & 59   & 30         \\
13-14            & --    & 90   & 47         \\
14-15            & 164    & 141  & 81     \\
15-16            & --    & 228  & 147        \\

\hline\hline
\end{tabular}
\end{center}
\tablecomments{These photometric precision measurements represent (in parts per million) the median 6 hour precision (as defined by VJ14) of all dwarf stars observed by K2 during the engineering test (ET) and during Campaign 0 (C0). \Kepler's photometric precision was calculated from the PDCSAP\_FLUX data in the \Kepler\ light curve files .}
\label{precisioncomp}

\end{table}

\subsection{Reflections of Jupiter}
\begin{figure*}[t!]
\epsscale{1}
  \begin{center}
      \leavevmode
\plotone{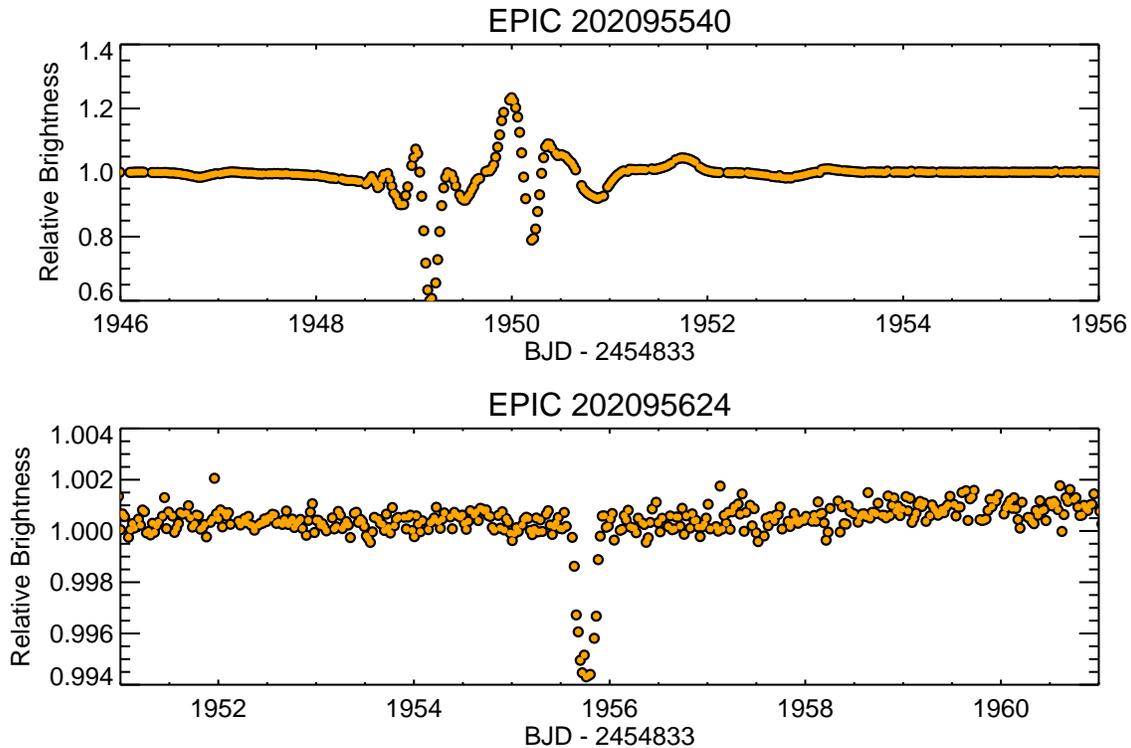}
\caption{Light curves contaminated by Jupiter's reflections. Top: Light curve from a Kp = 13.7 star on Module 23, Channel 79. A reflection of Jupiter passed over the target at about t = 1950, introducing significant artifacts into the light curve. Bottom: Light curve from a Kp = 13.7 star on Module 15 Channel 52.  The reflection of Jupiter leaving \Kepler's focal plane caused a spike in background flux, which left an artifact after subtraction of the median background flux from the aperture. This type of artifact can be confused for a transiting planet, but can easily be distinguished by the time of the event and the behavior of the background pixels at that time.\\ \\} \label{jupiter}
\end{center}
\end{figure*}
During Campaign 0, Jupiter fell in the field of view, in particular on Module 3 (which has been inoperable since the original \Kepler\ mission). Although Jupiter did not fall on active silicon, reflections and scattered light from Jupiter affect some of the light curves presented herein. For some stars, particularly those on or near Module 23, Channel 79, there are large variations in the background flux, which are not entirely removed during aperture photometry. An example of a light curve from this module is shown on the top in Figure \ref{jupiter}. 

Reflections from Jupiter also introduced artifacts in other modules when Jupiter exited the focal plane at BJD - 2454833 = 1955.75. There was a spike in the background level in many modules, and in some cases, the spike was not properly removed during aperture photometry. This can lead to anomalous spikes or decrements in the light curves at this time. An example of this effect in a light curve is shown on the bottom in Figure \ref{jupiter}. The decrements in particular are easy to mistake for transit events, so it is important to check the background for spikes when single transit-like events are detected. 

\section{Data Products}
\begin{figure*}[t!]
\epsscale{1}
  \begin{center}
      \leavevmode
\plotone{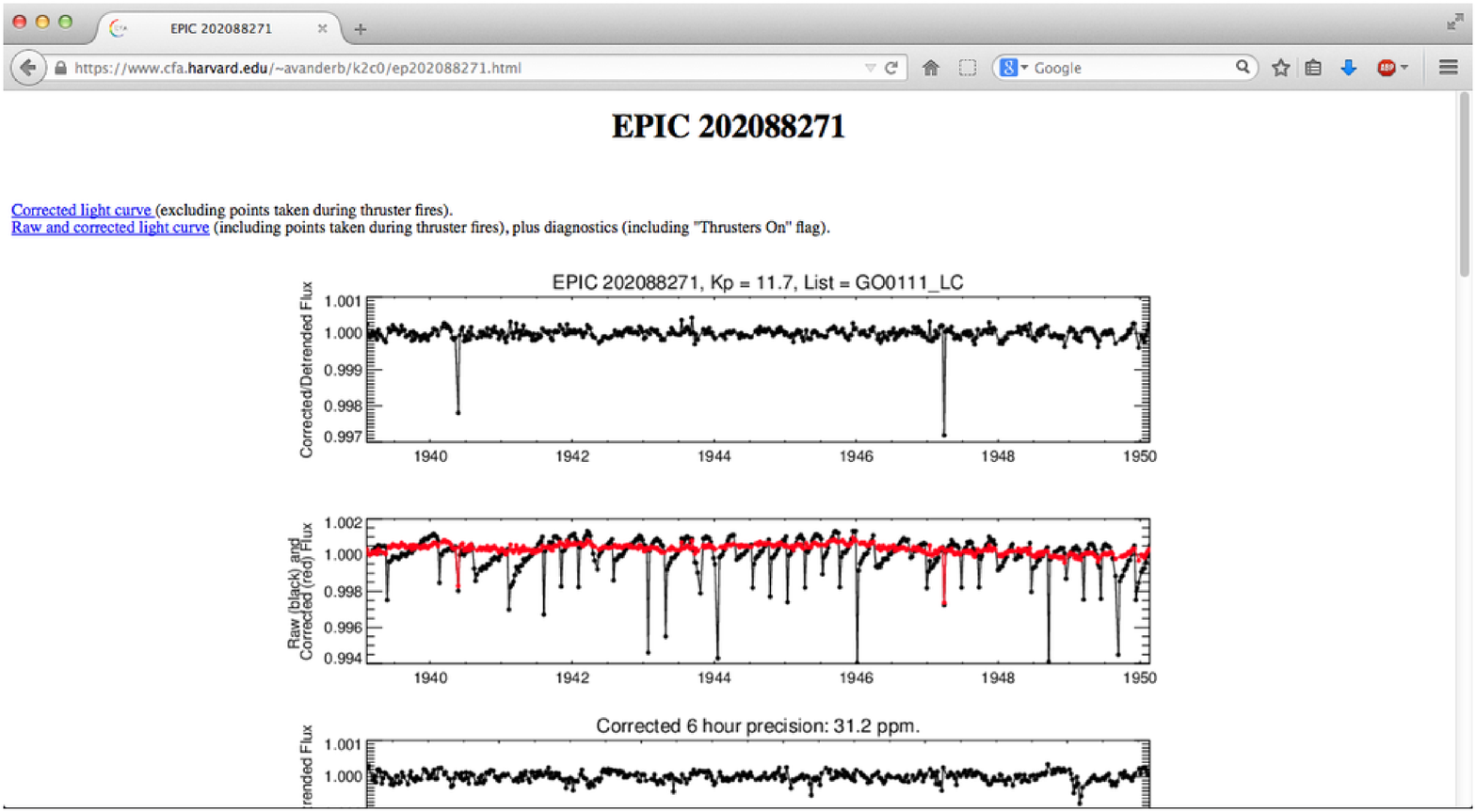}
\plotone{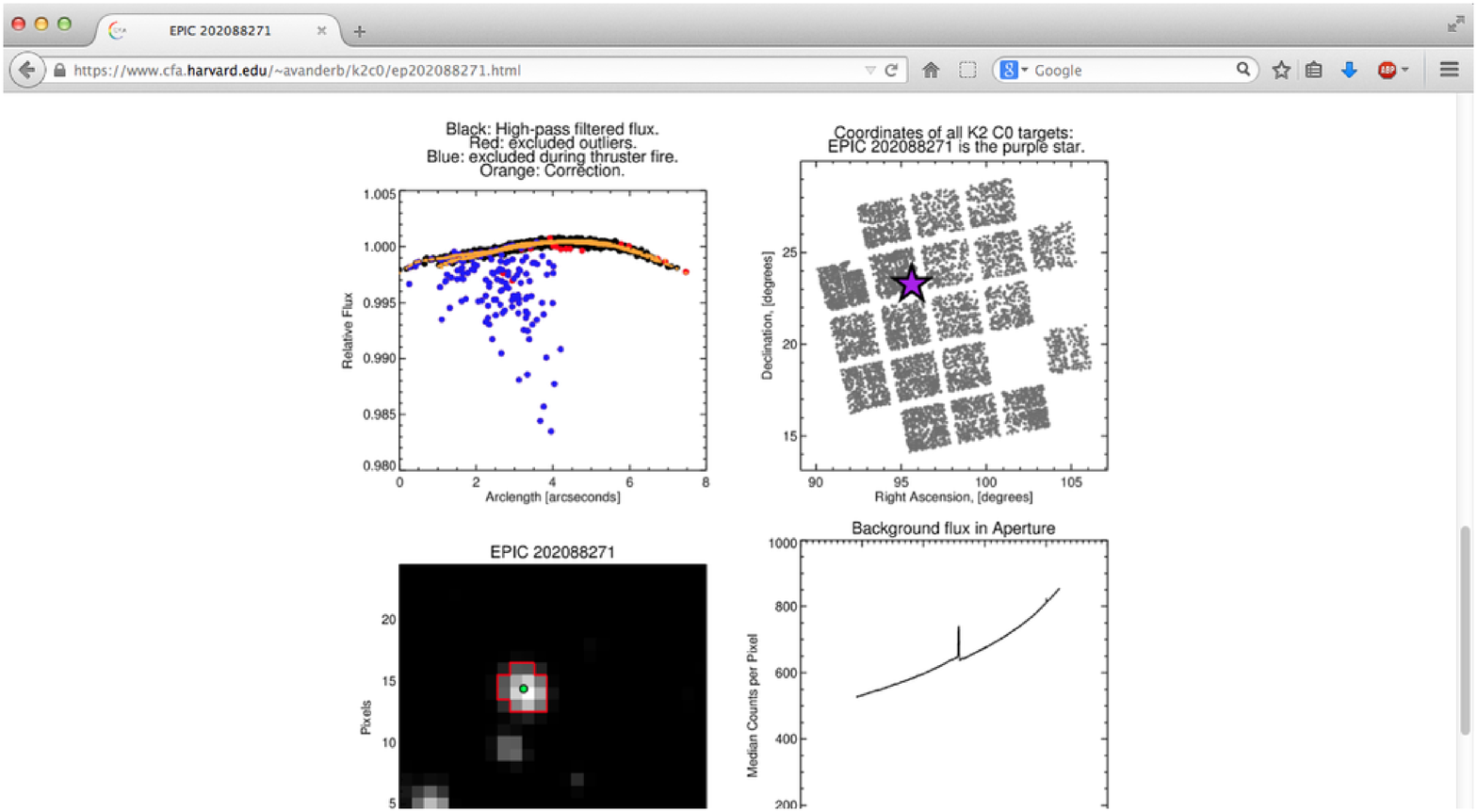}
\caption{Top: Screenshot of part of a webpage for an individual target showing links to light curve files and a light curve plot. Bottom: Screenshot of the same webpage showing diagnostic plots. } \label{screenshots}
\end{center}
\end{figure*}

We have made our reduced light curves available to the community online at the following URL: \url{https://www.cfa.harvard.edu/~avanderb/k2.html}. From this page, it is possible to download compressed files with all of our reduced light curves. We make our light curves available in two formats. The first is a simple format which is a two-column comma separated value (CSV) file with only the time of the observation and the SFF corrected light curve. In the simple files, thruster firing events have been automatically excluded. We also package our light curves in CSV files with columns for time, raw uncorrected flux, corrected flux, arclength (a measure of image centroid position as defined in VJ14), the measured flat field correction, and a flag indicating thruster fires. The additional information in the larger files makes it possible for users to re-derive the SFF correction under different assumptions and conditions. 

Additionally, we provide an online interface to quickly view and download individual light curves. We maintain a list of all of the Guest Observer targets observed during Campaign 0 at the following URL: \url{https://www.cfa.harvard.edu/~avanderb/allk2c0obs.html}. This webpage links to webpages for each individual target. Each target webpage includes links to the light curve CSV files for that particular object, as well as a plots of the raw and corrected light curve, and diagnostic information, including a plot showing the SFF correction, the background flux in the target pixels, the position of the target in relation to the rest of the K2 Campaign 0 targets, and an image of the target pixels and the photometric aperture. Screenshots of the webpage for one particular target are shown in Figure \ref{screenshots}. 

\section{Summary}

We have extracted photometric light curves from data taken by the \Kepler\ spacecraft during Campaign 0 of the K2 mission, corrected the light curves for the motion of the spacecraft using the technique of VJ14, and released the light curves to the community. These light curves have slightly worse photometric precision than measured during the K2 engineering test, possibly due to the fact that Field 0 is close to the plane of the Milky Way, while the engineering test field pointed out of the galaxy. We make our data available for download, and provide a simple interface for quickly looking at light curves online.

\acknowledgments
We acknowledge the tremendous effort of the K2 team and Ball Aerospace to make the K2 mission a success.  Some/all of the data presented in this paper were obtained from the Mikulski Archive for Space Telescopes (MAST). STScI is operated by the Association of Universities for Research in Astronomy, Inc., under NASA contract NAS5--26555. Support for MAST for non--HST data is provided by the NASA Office of Space Science via grant NNX13AC07G and by other grants and contracts. This paper includes data collected by the \Kepler\ mission. Funding for the \Kepler\ mission is provided by the NASA Science Mission directorate. This research has made use of NASA's Astrophysics Data System and the NASA Exoplanet Archive, which is operated by the California Institute of Technology, under contract with the National Aeronautics and Space Administration under the Exoplanet Exploration Program. A.V. is supported by the NSF Graduate Research Fellowship, Grant No. DGE 1144152.

Facilities: \facility{Kepler}

%\bibliographystyle{apj}
%\bibliography{refs}

\end{document}